\documentclass[useAMS,usenatbib]{mn2e}

\newcommand\ionm[2]{#1$\;${\small\rmfamily{#2}}\relax}
\newcommand{\be}{\begin{enumerate}}
\newcommand{\ee}{\end{enumerate}}

\newcommand{\msun}{\ifmmode {\rm M}_{\odot} \else M$_{\odot}$\fi}
\newcommand{\kms}{\ifmmode \mbox{\,km\,s}^{-1} \else \,km\,s$^{-1}$\fi}
\newcommand{\msfr}{\dot{M}_{\rm{SFR}}}
\newcommand{\mstar}{M_{\star}}
\newcommand{\mstarnot}{M_{\star,0}}

\newcommand{\oh}{[O/H]}

\newcommand{\tlogoh}{$12+\log(\mbox{O}/\mbox{H})$}

\newcommand{\zg}{Z_{\rm g}}
\newcommand{\zstar}{Z_{\star}}
\newcommand{\zsun}{Z_{\odot}}
\newcommand{\alphastar}{\alpha_{\star}}

\newcommand{\mutt}{\mu_{0.32}}

\newcommand{\mzr}{mass-metallicity relation}

\newcommand{\apj}{ApJ}
\newcommand{\apjl}{ApJL}
\newcommand{\apjs}{ApJS}
\newcommand{\aj}{AJ}

\newcommand{\mnras}{MNRAS}

\newcommand{\pasj}{PASJ}
\newcommand{\pasp}{PASP}
\newcommand{\araa}{ARA\&A}
\newcommand{\aap}{A\&A}

\usepackage{graphicx}
\usepackage{amssymb}
\usepackage{verbatim}
\usepackage{times}
\usepackage[usenames,dvipsnames]{color}
\definecolor{orange}{RGB}{255,127,0}
\begin{document}
\title[Stellar Metallicity Distributions]{An empirical prediction for
  stellar metallicity distributions in nearby galaxies}

\author[Peeples \& Somerville]{Molly S.\ Peeples$^{1}$\thanks{E-mail:
    molly@astro.ucla.edu} and Rachel S.\ Somerville$^{2}$
\\
$^{1}$Southern California Center for Galaxy Evolution Fellow, University of
California Los Angeles, 430 Portola Plaza, Los~Angeles,~CA~90095\\
$^{2}$Department of Physics and Astronomy, Rutgers, The State University of New Jersey, 136 Frelinghuysen Road, Piscataway,~NJ~08854
}

\pagerange{\pageref{firstpage}--\pageref{lastpage}} \pubyear{2012}

\date{\today}

\maketitle

\label{firstpage}

\begin{abstract}
  We combine star-formation histories derived from observations of
  high redshift galaxies with measurements of the $z\sim 0$ relation
  between gas-phase metallicity, stellar mass, and star formation rate
  to make an explicit and completely empirical connection between
  near-field and distant galaxy observations. Our approach relies on
  two basic assumptions: 1) galaxies' average paths through time in
  stellar mass vs.\ star formation rate space are represented by a
  family of smooth functions that are determined by the galaxies'
  final stellar mass, and 2) galaxies grow and become enriched with
  heavy elements such that they always evolve along the
  mass--metallicity--star formation rate relation.  By integrating
  over these paths, we can track the chemical evolution of stars in
  galaxies in a model independent way, without the need for explicit
  assumptions about gas inflow, outflow, or star formation efficiency.
  Using this approach, we present predictions of stellar metallicity
  (i.e., O/H) distribution functions for present day star-forming
  galaxies of different stellar masses and the evolution of the
  $\alpha$-element stellar metallicity-mass relation since $z\sim 1$.
  The metallicity distribution functions are fairly well described as
  Gaussians, truncated at high metallicity, with power-law tails to
  low metallicity. We find that the stellar metallicity distribution
  for Milky Way mass galaxies is in reasonable agreement with
  observations for our Galaxy, and that the predicted stellar mass
  vs. mean stellar metallicity relation at $z=0$ agrees quite well
  with results derived from galaxy surveys. This validates the
  assumptions that are implicit in our simple approach. Upcoming
  observations will further test these assumptions and their range of
  validity, by measuring the mean stellar mass-metallicity relation up
  to $z\sim 1$, and by measuring the stellar metallicity distributions
  over a range of galaxy masses.
\end{abstract}


\section{Introduction}\label{sec:intro}
As stars are born with the imprint of the elemental abundances in
their gaseous birth environments, measurements of stellar
metallicities within galaxies can be used to gain insight into their
star formation and chemical histories
\citep{tinsley75,tinsley78,blandhawthorn10}. In this paper, we carry
out an explicit test of the self-consistency between empirically
derived star formation histories, observational scaling relations
between gas phase metallicity ($\zg$), stellar mass ($\mstar$), and
star formation rate ($\msfr$), and observed stellar metallicities at
$z=0$. 

It has long been known that there is a strong positive correlation
between the luminosity or stellar mass of star forming galaxies and
their galaxy-averaged gas-phase abundance
\citep{garnett87,zaritsky94}.  This relationship (the ``\mzr'') has
now been measured for large samples of nearby galaxies
\citep[e.g.][]{tremonti04,kobulnicky04,kewley08} and for smaller
samples of distant galaxies out to $z\sim 3.5$
\citep{savaglio05,shapley05,erb06a,maiolino08,mannucci09}.  The
normalization, slope, scatter, and evolution of the observed \mzr\
place strong constraints on theoretical models of galaxy formation, in
particular on the physics of feedback from massive stars and
supernovae. Recently, it has been suggested that the scatter in the
observed \mzr\ can be reduced by considering the star formation rate
as a third parameter \citep{laralopez10,mannucci10}.  Physically, the
star formation rate is probably acting as a proxy for the galaxy gas
fractions, as galaxies with higher gas fractions have both higher star
formation rates and more diluted metals \citep{hughes12}.  Moreover,
\citeauthor{laralopez10} and \citeauthor{mannucci10} find that
galaxies up to $z\sim2.5$ \citep[e.g.,][]{savaglio05,erb06a} appear to
always lie on the same $\mstar$-$\zg$-$\msfr$ ``fundamental''
relation, though this may begin to break down at $z\sim 3$
(\citealp{maiolino08}, though see also \citealp{yabe12}).  It is not
yet clear whether or why a non-evolving
mass--metallicity--star-formation rate relation should be generically
expected, although some cosmological models do predict this
(\citealp{dave12}; M.\ Arrigoni et al.\ in
preparation). Self-regulation due to stellar-driven outflows is
usually invoked in this context
\citep{finlator08,spitoni10,peeples11,dave11,dayal12}.

There is a long history in the literature of attempts to model the
chemical evolution of galaxies using a ``classical'' approach, the
simplest version of which is the ``closed box'', in which galaxies
start out with all of their mass in the form of pristine gas, and
convert some fraction of that gas to stars, producing a certain
``yield'' of heavy elements along the way. It was quickly realized
that the simplest closed box picture could not reproduce observations,
in particular of the distribution of stellar metallicities in the
Solar Neighborhood in our Galaxy (the ``G-dwarf'' problem), leading to
modifications such as inflowing pristine or pre-enriched gas as well
as outflows of gas and metals
\citep{lyndenbell75,pagel89,colavitti08}. While much has been learned
from this approach, it has the obvious drawback that the results tend
to be dependent on the rather arbitrary functions adopted for the
inflow and outflow rates.

Another approach that has become fairly widely used is the coupling of
detailed chemical evolution models with cosmological models of galaxy
formation, realized in the form of either numerical hydrodynamic
simulations
\citep{scannapieco06,kobayashi07,brooks07,mouhcine08,oppenheimer08,wiersma09}
or semi-analytic models
\citep{delucia04,nagashima05,somerville08,arrigoni10}. These have the
advantage that quantities such as the rate of inflow of gas into
galaxies is motivated by the formation rate of structure in the Cold
Dark Matter (CDM) paradigm that provides the backbone for these
models, and outflow rates are also governed by physically motivated
scalings. However, they suffer from the drawback that currently, all
CDM-based galaxy formation simulations, whether numerical or
semi-analytic, apparently fail to reproduce the observed scaling of
star formation history functional form with galaxy mass (sometimes
referred to as ``downsizing''; see \citealp{fontanot09} for a detailed
discussion).

Here we propose a different approach to modeling the buildup of metals
in galaxies that sidesteps some of the problems with both the
classical and cosmological methods. We make use of a family of
parameterized empirical star formation histories, derived from
observations of the stellar mass vs.\ star formation rate relation at
different redshifts by \citet{leitner12}. These empirically-derived
star formation histories are in good agreement with those derived from
spectral energy distribution modeling and the evolution of the
stellar-halo mass relation
\citep{zheng07,conroy09a,behroozi12,moster12}.  We assume that the
{\em average} star formation history for galaxies with a given stellar
mass today can be represented by this smooth functional form. This
assumption is supported by mounting observational evidence that
galaxies with masses of roughly the Milky Way or lower ($\sim$ few
$\times 10^{10}\, \msun$) build up most of their mass through smooth
accretion of gas, with mergers and starbursts playing a relatively
minor role
\citep{noeske07a,robaina09,oliver10,karim11,leitner12}. This is also
supported by results from cosmological simulations
\citep{brooks09,hirschmann12,tissera12}. Then, once stellar mass loss
and recycling has been accurately accounted for \citep{leitner11}, the
growth of galaxies can be traced through cosmic time by stepping along
these $\msfr(\mstar)$ relations as a function of redshift.

We make a second assumption, that galaxies always evolve {\em along}
the $\mstar$-$\zg$-$\msfr$ relation. That means that as galaxies
evolve along their trajectories in $\mstar$-$\msfr$ space (specified
by their star formation history), we assume that the stars born at
that moment form out of gas with a metallicity $\zg$ given by the
$\mstar$-$\zg$-$\msfr$ relation. In this way, we can build up the
{\em distribution function} of stellar metallicities for galaxies with
different masses at $z=0$. We can also test whether the two
assumptions of empirical star formation histories plus a non-evolving
$\mstar$-$\zg$-$\msfr$ relation lead to an average stellar metallicity
relation at $z=0$ that agrees with observations. Both of these
quantities will be better constrained observationally over a wider
range of galaxy masses, and to higher redshift, in the near future.

This paper is organized as follows.  In \S\,\ref{sec:relations}, we
describe our adopted and predicted scaling relations connecting galaxy
stellar masses, star formation rates, and gas-phase metallicities
across cosmic time.  In \S\,\ref{sec:stars} we present the stellar
metallicity distribution functions and evolution of the stellar
metallicity-mass relation predicted by this model.  We conclude with a
brief discussion of relevant implications in \S\,\ref{sec:conc}.

Throughout we assume a \citet{chabrier03b} stellar initial mass
function (integrated over $0.1$--$100\,\msun$) and \tlogoh$_{\odot}=8.7$
\citep{asplund09}, although we note that the 8.9 value found from
helioseismology \citep{delahaye06} would lead to $0.2$\,dex lower
$Z/\zsun$ values than stated here.  Following \citet{leitner12}, we
adopt a flat $\Lambda$CDM cosmology with $\Omega_{\rm
  M}=1-\Omega_{\Lambda}=0.258$ and $h=0.72$.

\section{Evolution of Scaling Relations}\label{sec:relations}
The model we present here is based on two simple assumptions: 1)
galaxies' average star formation histories are given by the empirical
functions provided by \citet{leitner12}, in which their $z=0$ stellar
mass determines the shape of their star formation histories. 2)
Galaxies evolve along a ``fundamental'' $\mstar$-$\zg$-$\msfr$
relation that is itself unchanging with redshift.  Taken together,
these pieces provide a unique enrichment history for an average galaxy
of a given present day stellar mass $\mstarnot$. In this section we
describe these two pieces of our model, where they come from, and some
caveats, in a bit more detail.

\begin{figure*}
\includegraphics[width=0.48\textwidth]{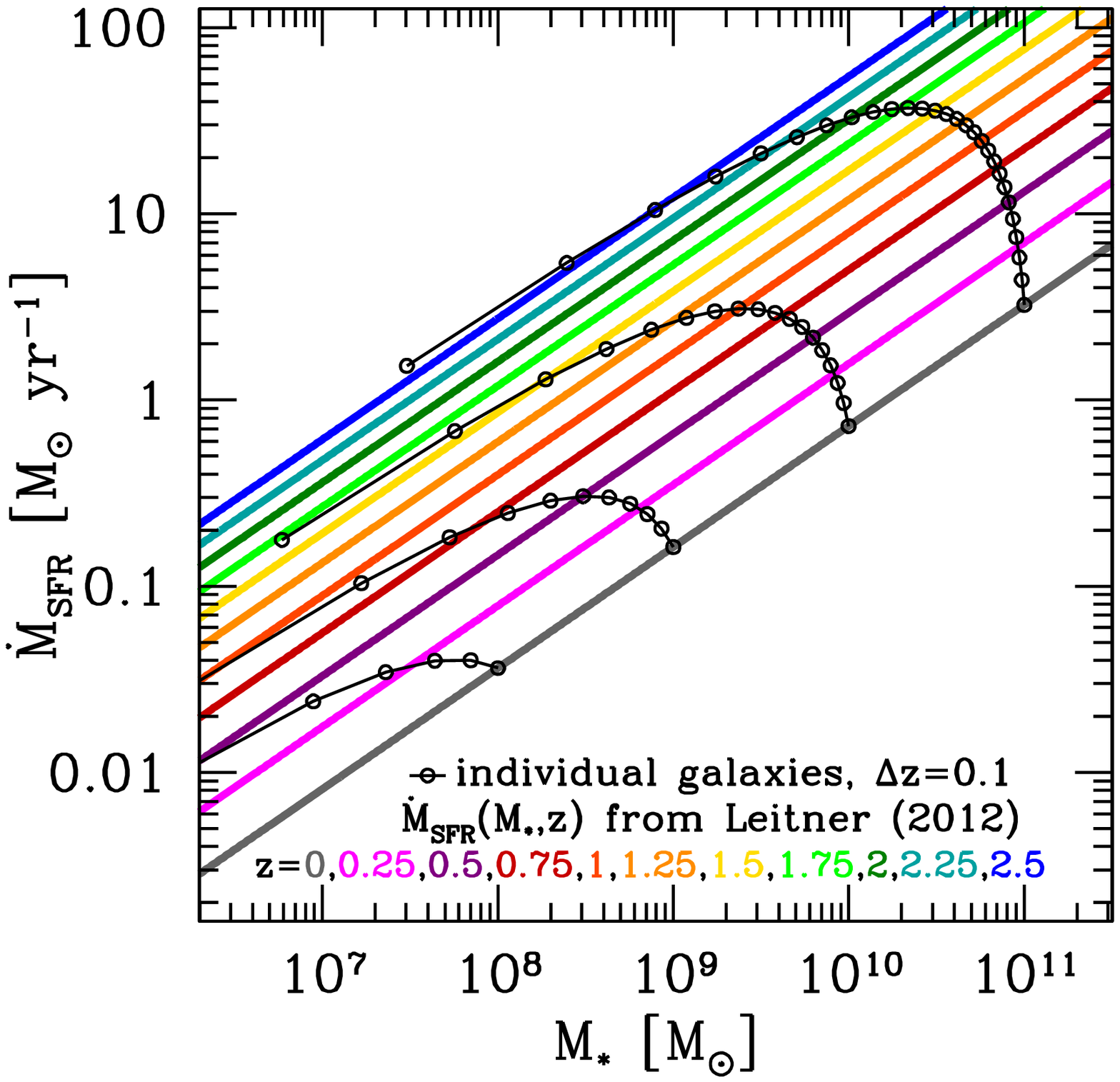}\hfill
\includegraphics[width=0.48\textwidth]{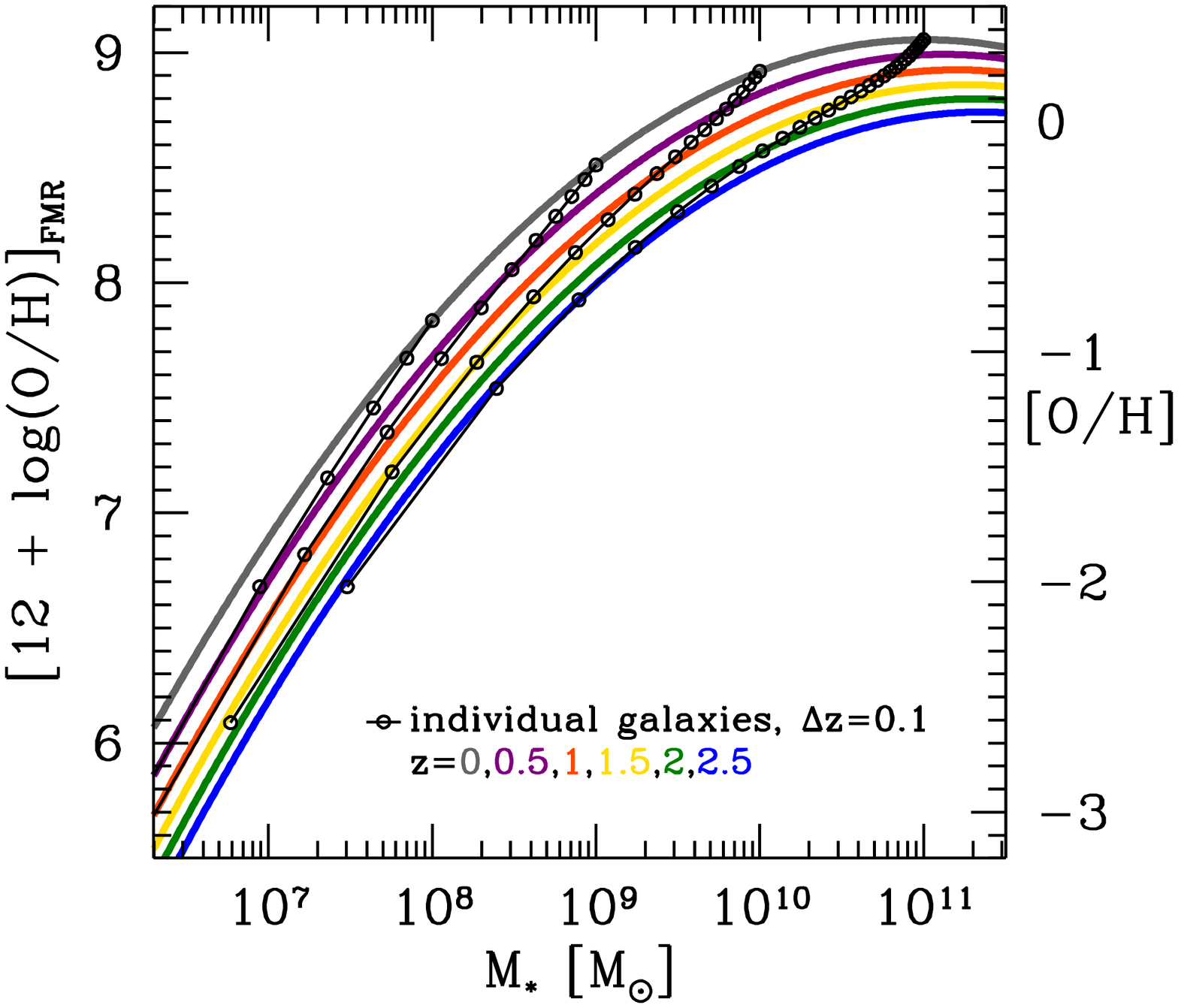}\\
\includegraphics[width=0.48\textwidth]{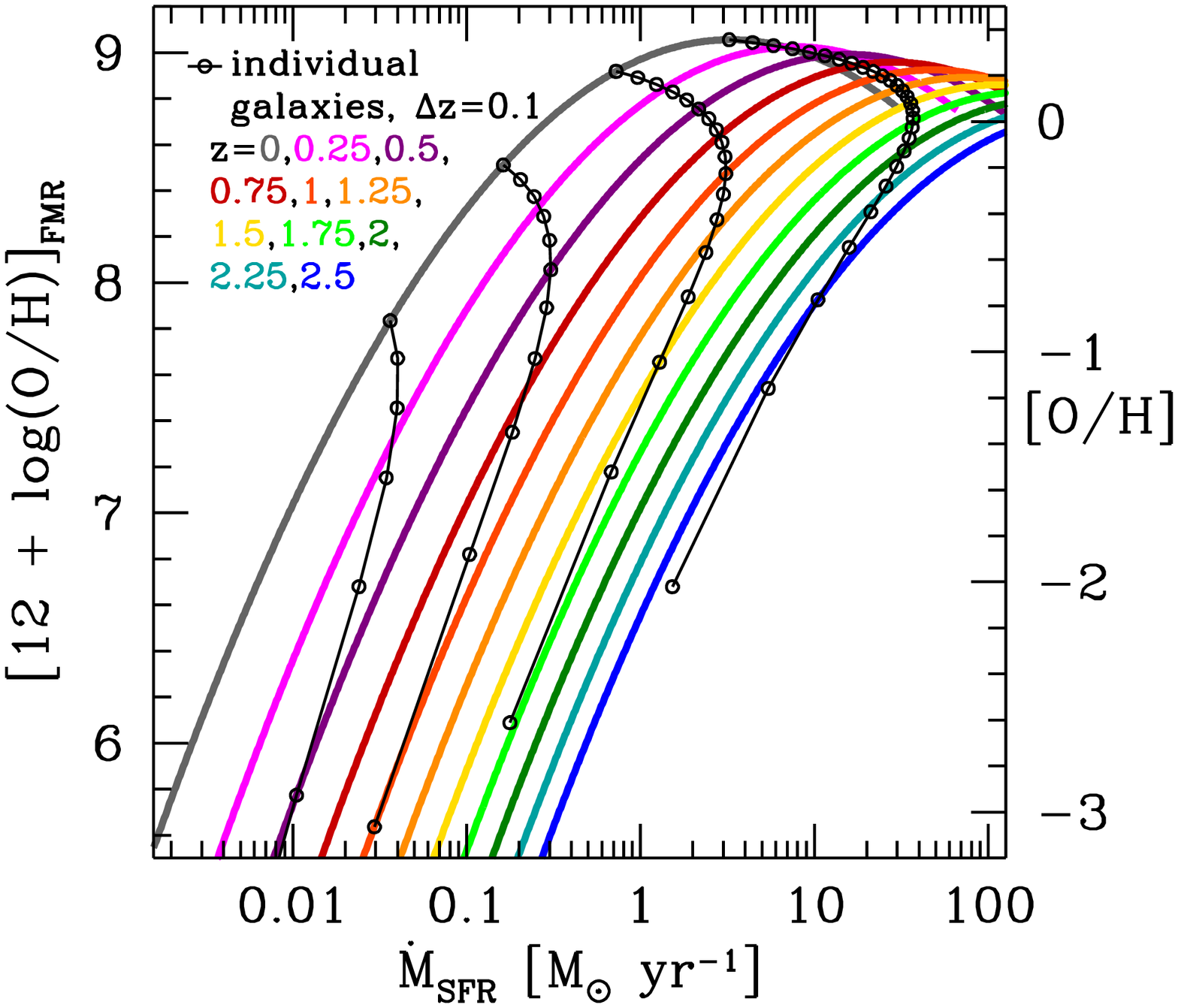}\hfill
\includegraphics[width=0.48\textwidth]{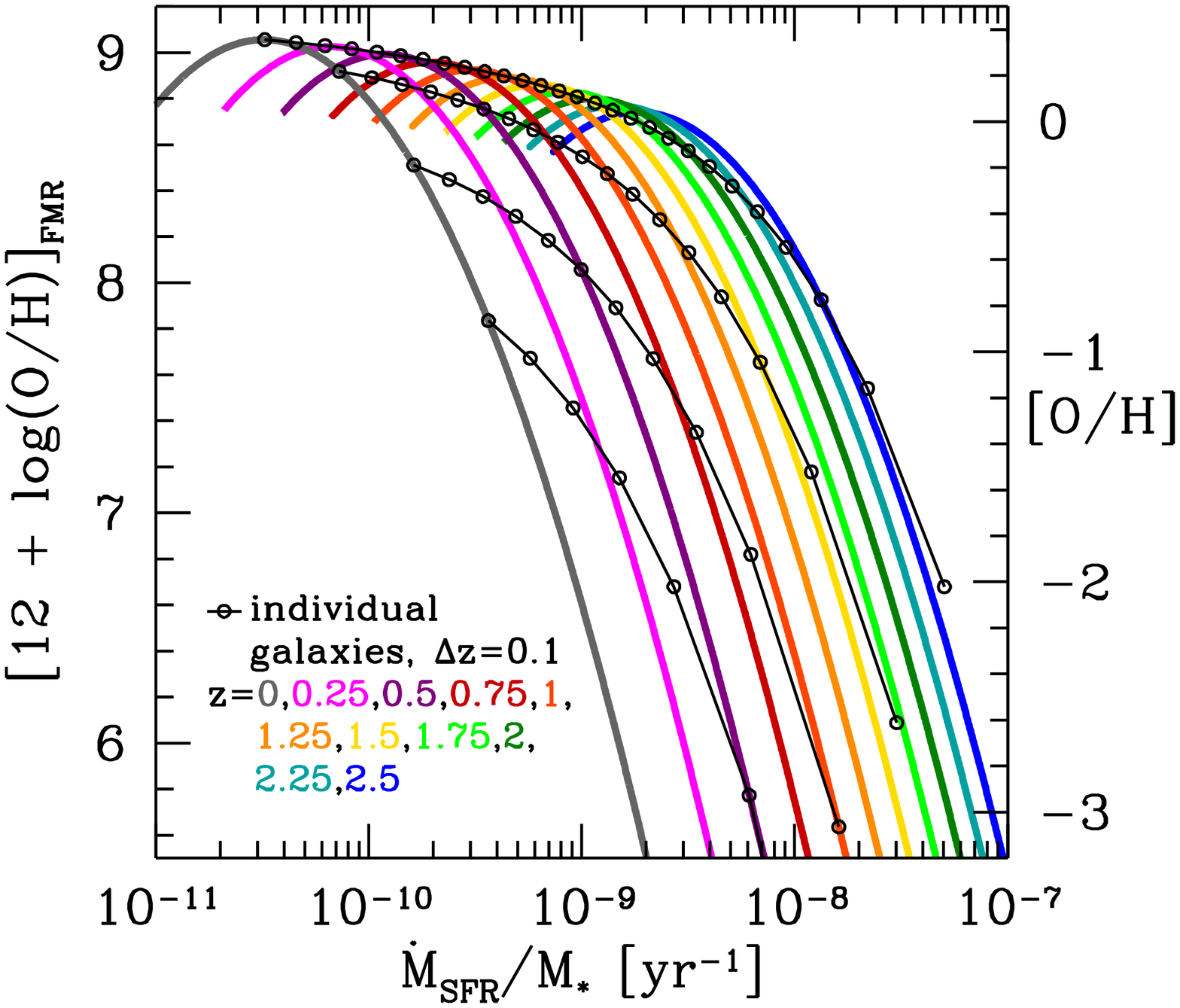}
\caption{\label{fig:scaling} Evolution of the $\mstar$-$\msfr$
  relation as a function of redshift.  Colored lines show
  ``snapshots'' of the galaxy population at different redshifts;
  growth of individual galaxies is shown by black curves with
  dots. {\em Top right:} Evolution of the \mzr\ as predicted by the
  $\mstar$-$\zg$-$\msfr$\ relation. {\em Bottom left:} Evolution of
  the $\zg$-$\msfr$ relation.  {\em Bottom right:} Evolution of $\zg$
  as a function of the specific star formation rate, $\msfr/\mstar$.
  In all panels, individual galaxies are shown in black, from left to
  right, with $\log\mstar/\msun=8,\,9,\,10,\,11$ at $z=0$. }
\end{figure*}

The starting point for the first piece of the model is measurements of
galaxy stellar masses and star formation rates from $z\sim 0$ to 2.5 by
\citet{karim11}, from the COSMOS survey. It is now well established
that $\mstar$ and $\msfr$ show a fairly tight and nearly linear
relationship which evolves in normalization, but very little in slope,
over a very broad redshift range \citep{noeske07a,daddi07,karim11,elbaz11}.
These results of course represent snapshots of different galaxy
populations at different cosmic times.  \citet{leitner12} first fit
power laws to these snapshots in time, using the $\msfr(\mstar,z)$
results measured by \citet{karim11},
\begin{equation}\label{eqn:sfms}
\msfr(\mstar,z) = A_{11}\left(\frac{\mstar}{10^{11}\,\msun}\right)^{\beta+1}(1+z)^{\alpha},
\end{equation}
where $A_{11}=0.0324\times10^{11}\,\msun$\,Gyr$^{-1}$, $\alpha=3.45$,
and $\beta=-0.35$. He then characterized how {\em individual}
galaxies would have to grow over time in $\mstar$ vs. $\msfr$ such
that the evolution of these relations would be satisfied, accounting
for mass loss and recycling from evolving stellar populations using
the method presented in \citet{leitner11}. Thus, these
$\msfr(\mstar,z)$ relations can be integrated to give the
star-formation history for a galaxy of a given stellar mass
$\mstarnot$ at $z=0$, $\msfr(\mstarnot,z)$\footnote{A similar
  approach, with qualitatively similar results, has been presented by
  \citealp{noeske07a}}. The top left panel of Figure\,\ref{fig:scaling}
shows the fits representing the population averages at different
redshifts, with the trajectories of individual galaxies of different
$\mstarnot$ overplotted.

The two obvious limitations of this method are that it only describes
galaxies that are actively forming stars at $z=0$ (i.e., there is no
accounting for a galaxies that cease forming stars and ``fall off'' of
these relations) and that it is only valid up to redshifts and stellar
masses where the $\mstar$-$\msfr$ relation is reliably measured.
However, the majority of galaxies with stellar masses in the range
$10^9<\mstar< 3 \times 10^{10}\,\msun$ are on the ``star forming main
sequence'' \citep{brinchmann04,salim07,kimm09}, i.e. on the relations
used to derive the \citet{leitner12} star formation histories. Based
on observational estimates of merger rates \citep{robaina09}, the
scatter in the observed $\mstar$-$\msfr$ relation \citep{noeske07a},
and theoretical predictions \citep{brooks09}, galaxies in this mass
range are expected to build up the majority of their mass through
smooth accretion, with mergers and starbursts playing a minor role.
On the second point, the star formation histories
\citeauthor{leitner12}\ derives from \citeauthor{karim11}'s data are
not entirely reliable at $z\gtrsim1.6$ due to observational
uncertainties (i.e., the redshifts for which $\msfr$ has not been
well-measured for $\mstar\lesssim 10^{9.6}\,\msun$).  We further note
that the star formation histories for $\mstar=10^8\,\msun$ galaxies at
$z=0$ are entirely based on extrapolations, as the progenitors of
these galaxies cannot currently be observed at higher redshift. This
is true to some extent at somewhat higher masses as well
($10^8<\mstar< 3 \times 10^{9}\,\msun$), with extrapolation becoming
less important as $\mstarnot$ increases.  \citeauthor{leitner12}\
notes that the star formation histories derived for $\mstar\sim
10^8\,\msun$ galaxies predict a much larger fraction of recent star
formation than those derived from resolved stellar population studies:
the \citeauthor{leitner12}\ star formation histories for these low
mass galaxies imply that all of their star formation takes place at
$z<1$, while the resolved stellar population studies indicate that a
significant fraction of past star formation occurred at $z\gg 2$
\citep{weisz11}.

The other three panels show the predicted evolution of the
$\zg$-$\mstar$, $\zg$-$\msfr$, and $\zg$-$\mstar/\msfr$ relations,
where for the gas-phase metallicity we plot the observable \tlogoh\ as
would be measured from star-forming \ionm{H}{II}\ regions derived from
the $z\sim 0$ $\mstar$-$\zg$-$\msfr$ relation.  While several fits to
the $\mstar$-$\zg$-$\msfr$\ relation exist in the literature, we
choose to calculate metallicities from the double-quadratic fit
\begin{equation}\label{eqn:fmr}
\begin{array}{l}
[12+\log(\mbox{O}/\mbox{H})]_{\mbox{\scriptsize FMR}} =\vspace*{0.5em}\\
\qquad 8.90 + 0.37m- 0.14s - 0.19m^2+ 0.12ms - 0.054s^2,
\end{array}
\end{equation}
where $m=\log\mstar-10$ and $s=\log\msfr$, fit to Sloan Digital Sky
Survey (SDSS) data by \citet{mannucci10}. \citeauthor{mannucci10}\
postulate that this so-called ``fundamental metallicity relation''
(FMR) holds at all redshifts, and show that the observed $z\lesssim
2.5$ galaxies fall along this relation when extrapolated to higher
star formation rates. Equation\,(\ref{eqn:fmr}) is fit over the ranges
$9.1\lesssim\log\mstar\lesssim11.35$ and $-1.45\lesssim\msfr\lesssim
0.80$ (see \citealt{mannucci10} for the exact $\msfr[\mstar]$ bins
used in their calculation); we must extrapolate outside of this range
in order to describe the metallicity evolution of
$8\leq\log\mstar/\msun\leq 11.5$ galaxies at $z=0$.  This
parameterization is not \citeauthor{mannucci10}'s best fit to the
data, but we find that this functional form is better suited to the
necessary extrapolation than alternatives.  Specifically,
Equation\,(\ref{eqn:fmr}) will under-predict the metallicities for
low-mass galaxies, compared to a more robust fit using
$\mutt\equiv\log\mstar-0.32\log\msfr$ as the independent variable,
\begin{equation}\label{eqn:mutt}
\begin{array}{l}
[12+\log(\mbox{O}/\mbox{H})]_{\mbox{\scriptsize FMR},\mu} =\vspace*{0.5em}\\
\qquad 8.90 + 0.39x - 0.20x^2 - 0.077x^3 + 0.064x^4,
\end{array}
\end{equation}
where $x\equiv\mutt-10$ \citep{mannucci10}.  Unfortunately, this
quartic parameterization leads to an up-turn in \tlogoh\ when
extrapolated to low $\mutt$; while no extrapolation in $\mutt$ from
$z=0$ to $z\sim2$ is needed to describe {\em observed} galaxies,
extrapolation {\em is} required to describe the evolution of the local
galaxies we wish to characterize.  Finally, \citet{laralopez10}
describe the relation between stellar mass, gas-phase metallicity, and
star formation rate as a plane,
\begin{equation}\label{eqn:lara}
\begin{array}{l}
[12+\log(\mbox{O}/\mbox{H})]_{\mbox{\scriptsize L10}} =\vspace*{0.5em}\\
\qquad 0.891\log\mstar - 0.422\log\msfr + 0.086.
\end{array}
\end{equation}
We choose, however, to not adopt this formulation because it gives
even lower metallicities at high-redshift than
Equation\,(\ref{eqn:fmr}), while overestimating metallicities at high
$\mstar$.

Finally, we note that uncertainties in the calibration methods for
determining \tlogoh\ compound the difficulty in accurately
characterizing the $\mstar$-$\zg$-$\msfr$\ relation
\citep[e.g.,][]{kewley08}; these uncertainties not only affect the
slope of the relation, but also its normalization at the $\sim
0.3$\,dex\ level.\footnote{Comparisons of gas-phase and stellar
  metallicities could provide an potentially useful avenue for placing
  combined constraints on ``reasonable'' ranges for the Solar oxygen
  abundance, normalization of the \mzr, and the nucleosynethetic
  oxygen yield.}  Thus it will certainly be possible to improve on
the results presented here once the $\mstar$-$\zg$-$\msfr$\ relation
has been measured to lower masses (e.g., B.\ Andrews \& P.\ Martini,
in preparation) and with more robust metallicity calibrations (e.g.,
\citealp{nicholls12}; \citealp{lopezsanchez12}; M.\ Dopita et al. in
preparation).

From the top-right panel of Figure\,\ref{fig:scaling}, it is obvious
that on short timescales, galaxies evolve along the mean \mzr\
\citep[c.f.,][]{peeples11,dave12}.  Though it is subtle, the \mzr\
evolves slightly more rapidly at low masses than at high masses. The
bottom two panels show that this is because massive galaxies
asymptotically plateau in their metallicities, while low-mass
galaxies are still rapidly evolving at late times.  The galaxy tracks
in the bottom-left panel of Figure\,\ref{fig:scaling} show the
metallicities at which galaxies are at the peak of their star
formation.  In \S\,\ref{sec:stars}, we integrate along these tracks to
derive stellar metallicities as a function of cosmic time.

\section{An Empirical Prediction for Stellar Metallicities}\label{sec:stars}
\begin{figure}
\centering
\includegraphics[height=0.26\textheight]{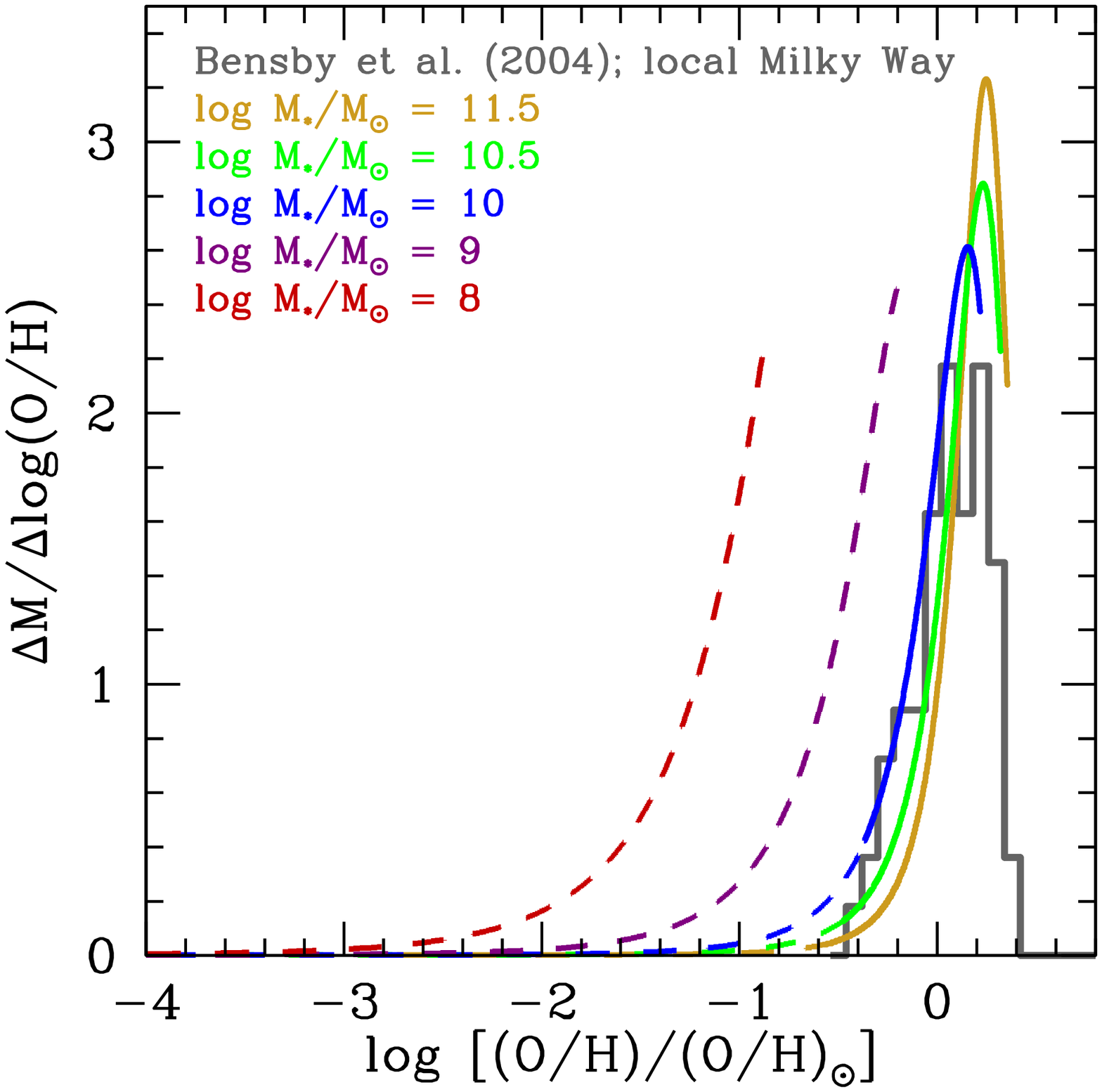}\\
\includegraphics[height=0.26\textheight]{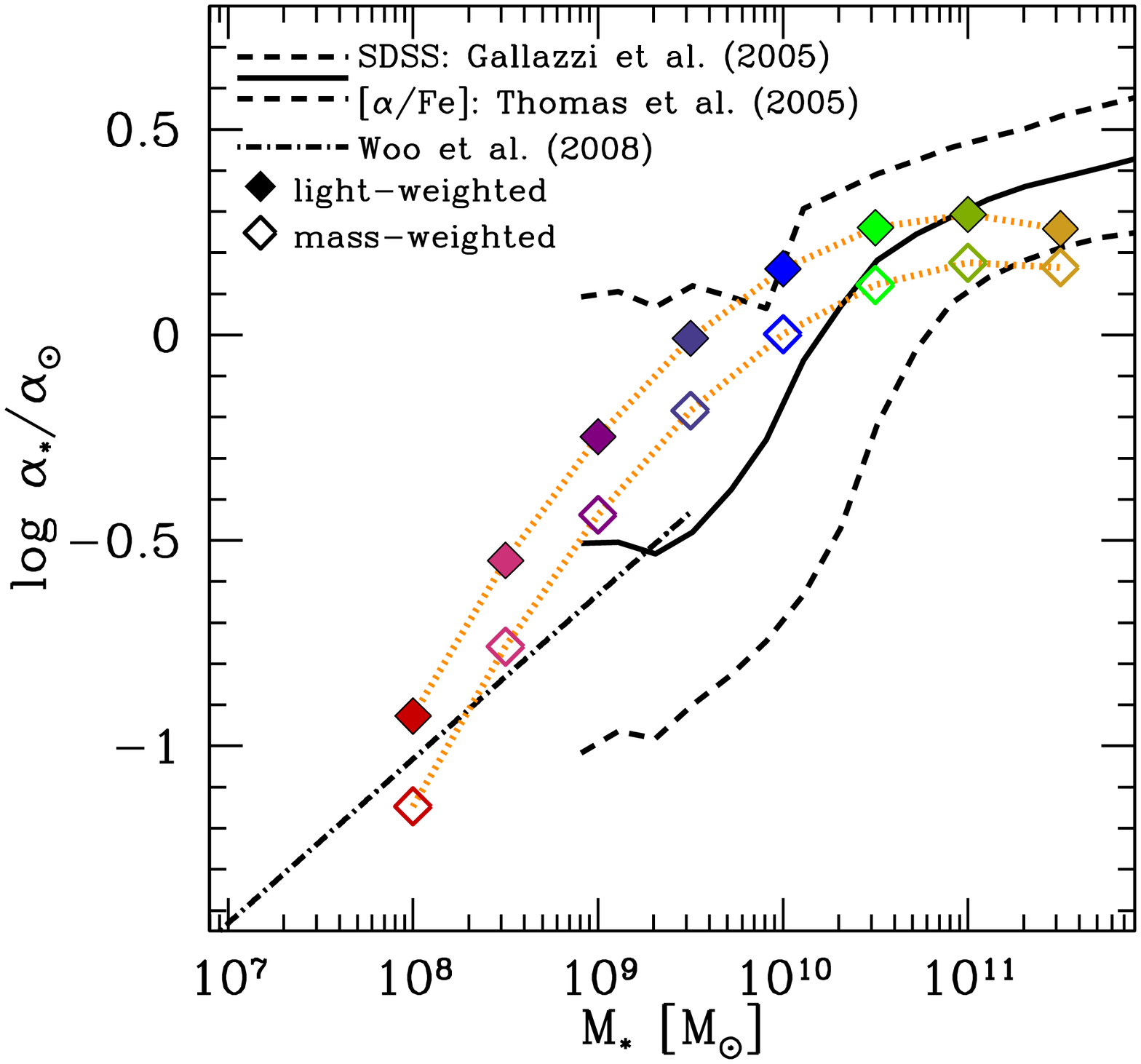}\\
\includegraphics[height=0.26\textheight]{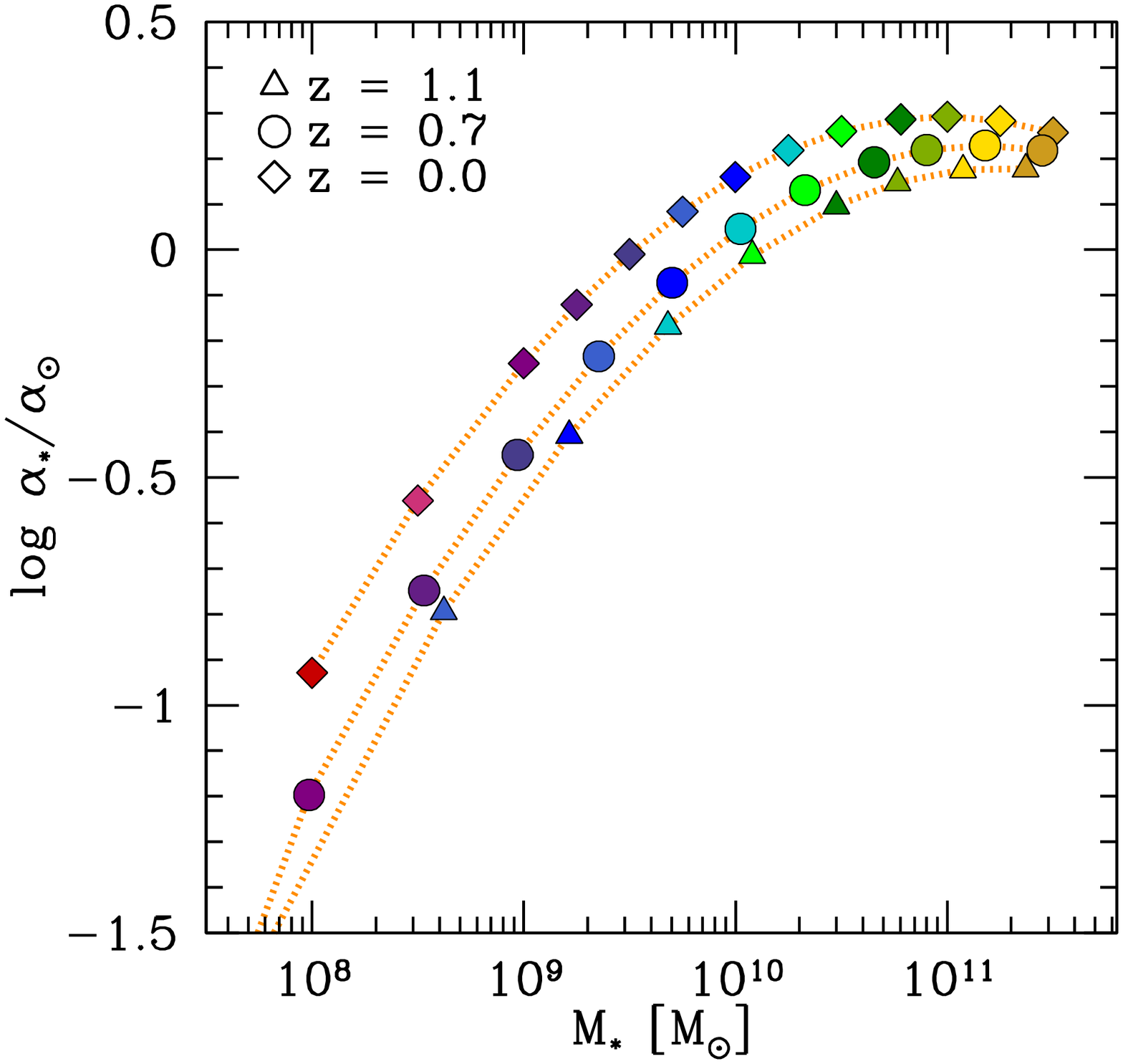}
\caption{\label{fig:stellar}{\em Top:} Distribution of stellar \oh\ in
  galaxies of different $\mstarnot$. Values are given in
  Table\,\ref{tbl:hist}. Measurements of the Milky Way disk from
  \citet{bensby04} are shown for reference. Dashed lines denote
  extrapolation. {\em Middle:} The stellar metallicity-mass relation;
  data from SDSS \citep{gallazzi05} is shown as the black solid line
  (dashed lines showing $\pm 1\sigma$ dispersion) using the
  $[\alpha/{\rm Fe}]$-$\mstar$ conversion from \citet{thomas05}. The
  fit to the $\zstar$-$\mstar$ relation for dwarfs by \citet{woo08} is
  shown as a dot-dashed line.  The open symbols denote mass-weighting,
  while the filled symbols are the $\alpha$ abundances weighted by the
  $B$-band luminosity. {\em Bottom:} The $\alphastar$-$\mstar$
  relation at $z=0$ ({\em diamonds}), 0.7 ({\em circles}), and 1.1
  ({\em triangles}).  Symbols of the same color denote tracks of
  individual galaxies.  Values are given in
  Table\,\ref{tbl:mzevolve}.}
\end{figure}

Because stars are born with the chemical imprint of the gas they
formed out of, by tracing galaxies through the evolutionary tracks
plotted in Figure\,\ref{fig:scaling}, we are able to track---in an
average sense---the buildup and change in stellar metallicities in
different galaxies.  Figure\,\ref{fig:stellar} shows the distribution
of the stellar [$\alpha$/H] within individual galaxies of different
$\mstar$ at $z=0$ and the predicted relation between galaxy-averaged
metallicities ($\zstar$) and $\mstar$.  Though the distribution of
stellar [Fe/H] has been well-measured for the Milky Way
\citep[e.g.,][]{ivezic08}, [$\alpha$/H] has not been as well
characterized.\footnote{In an effort to keep our model as
  empirically-based as possible, we do not try to model Iron
  production directly.}  The histograms in the top panel of
Figure\,\ref{fig:stellar} are based on {\em all} star formation, and
thus are only applicable for comparing to metallicity distributions of
low-mass stars, such as for the upcoming SDSS-III APO Galactic
Evolution Experiment (APOGEE) observations of the Milky Way
\citep{allende08,eisenstein11} and $\mstar\sim 10^8\,\msun$ dwarf
Irregular galaxies in the Local Group (E.\ Kirby et al.\ in
preparation; see also \citealp{kirby10,kirby11b}).  We note that our
assumption that all galaxies of a given present-day stellar mass
follow the same smooth star formation history may tend to
underestimate the width of the ensemble-averaged stellar metallicity
distribution function. It is well known that galaxies have somewhat
stochastic star formation histories (although apparently they
typically do not deviate from the star forming main sequence by more
than a factor of 2--3; \citealp{noeske07a}), and that the overall
shape of the star formation history can show considerable diversity
from galaxy to galaxy, particularly for low mass objects
\citep[e.g.,][]{weisz11}. These effects would tend to broaden the
predicted metallicity distribution function and to lead to a
corresponding galaxy-to-galaxy diversity in this quantity. However,
they should not significantly affect our predictions for the galaxy-
and ensemble-averaged mass metallicity scaling relations that we
discuss below.

The $\log\mstar/\msun=10.5$ (bright green) distribution is included
for comparison to the Milky Way \citep{klypin02}; we also show the
distribution of [O/H] as measured for 72 local disk stars by
\citep{bensby04}, though we caution that it is unclear the extent to
which these stars are representative of the entire Milky Way.  The
dashed portions denote where Equation\,(\ref{eqn:fmr}) has been
extrapolated.  At higher stellar masses, the metallicity distribution
functions turn over; this is directly related to the decline in star
formation rates at later times, resulting in the formation of
relatively few higher-metallicity stars (see the bottom-left panel of
Figure\,\ref{fig:scaling}).  Because the turnover in the star
formation histories is relatively slow, the peak in the \oh\
distribution function is at higher metallicity than the \oh\ at which
the galaxy is when at its peak star formation rate.  Since
$\mstarnot\sim 10^{11}\,\msun$ galaxies are skirting the peak of the
$\zg$-$\mstar/\msfr$ relation (bottom-right panel of
Figure\,\ref{fig:scaling}), they have the nearly maximal stellar
[$\alpha$/H] distribution; the turnover in the metallicity scaling
relations at high $\mstar$ and high $\msfr$ is why the
$\log\mstar/\msun=10.5$ and $11.5$ distribution functions are
surprisingly similar.

Table\,\ref{tbl:hist}\ gives the distribution functions plotted in the
top panel of Figure\,\ref{fig:stellar}. At $\log\mstar/\msun\lesssim
9.25$ (i.e., the regimes that are calculated entirely from
extrapolations), the distribution functions at are fairly well
described by power-laws of the form
\begin{equation}\label{eqn:alphapwr}
\log\frac{\Delta M}{\Delta\log({\rm O/H})} = a[{\rm O/H}] + b,
\end{equation}
where $\Delta M$ is the mass fraction in a bin of width
$\Delta\log({\rm O/H})$, $[{\rm O/H}]\equiv\log[({\rm O/H})/({\rm
  O/H})_{\odot}]$, and $a$ and $b$ are functions of stellar mass such
that
\begin{eqnarray}\label{eqn:alphafits}
a & =&  0.2205\log\mstar/\msun - 0.9305\quad\mbox{and}\\
b & =& -0.42\log\mstar/\msun + 4.36.
\end{eqnarray}
At higher masses, this power-law distribution turns over, such that
the high-metallicity distribution is well-described as a truncated
Gaussian, where the maximum metallicity is given by the $z=0$
gas-phase metallicity (top-right panel of Figure\,\ref{fig:scaling}).

\begin{table}
\centering
\begin{tabular}{rr}\hline\hline
$\log(\alphastar/\alpha_{\odot})$ & $\Delta M/\Delta\log(\mbox{O/H})$ \\\hline
$-3.3907$ & $-0.00006$ \\
$-2.9125$ & $-0.00011$ \\
$-2.5897$ & $-0.00024$ \\
$-2.3483$ & $-0.00042$ \\
$-2.1570$ & $-0.00066$ \\
$-1.9997$ & $-0.00097$ \\
\vdots & \vdots \\
$0.3202$ & $2.23054$ \\
$0.3203$ & $2.22968$ \\
$0.3205$ & $2.22867$ \\
$0.3206$ & $2.22769$ \\
$0.3208$ & $2.22358$ 
\end{tabular}
\caption{\label{tbl:hist} 
  Stellar metallicity distribution function for
  $\mstarnot=10^{10.5}\,\msun$ galaxy as plotted in the top panel of
  Figure\,\ref{fig:stellar}.  Negative  $\Delta M/\Delta\log(\mbox{O/H})$ denotes where Equation\,(\ref{eqn:fmr}) has been extrapolated.   Full tables for
  $\log\mstarnot/\msun=8,\,9,\,10,\,10.5,\,11$ and 11.5 are available in the electronic version.}
\end{table}

The bottom two panels of Figure\,\ref{fig:stellar} show the
galaxy-averaged $\alpha$-element stellar metallicities ($\alphastar$)
as a function of stellar mass at $z=0$ (middle panel) and at $z=0.7$
and $z=1.1$ (bottom panel).  The open symbols in the middle panel show
the stellar mass-weighted mean $\alphastar$-$\mstar$ relation.  Though
it is a small effect, mass loss must be taken into account to
correctly compute the mass-weighted average for an entire galaxy; we
use the methods of \citet{leitner11} for consistency with our adopted
star formation histories.  The solid symbols denote a $B$-band
luminosity weighted mean, which we obtain using the stellar population
models of \citet{bruzual03} and our assumed star formation
history. The $B$-band luminosity weighted mean metallicity closely
approximates observational estimates of metallicity from spectral
lines \citep{trager09}.  For comparison, we plot the SDSS stellar
metallicity-mass relation as measured by
\citet{gallazzi05}.\footnote{We note that our model predicts
  metallicities for stars formed {\em in situ}, whereas many
  star-forming galaxies have a bulge component that is thought to have
  been formed from the merging of smaller galaxies
  \citep[e.g.,][]{toomre77,kormendy04}.  Because the SDSS spectra are taken
  from fibers placed on the central 3\arcsec\ of each galaxy, the
  derived metallicities could potentially be highly sensitive to such
  a bulge component.} The \citeauthor{gallazzi05}\ measurements are
primarily sensitive to iron abundances (A.\ Gallazzi, private
communication), but our model is implicitly tied to oxygen abundances
(i.e., $\alpha$ elements). Therefore, we renormalize the
\citeauthor{gallazzi05}\ data using the [$\alpha$/Fe] ratios (relative
to Solar) found by \citet{thomas05}, where
\begin{equation}\label{eqn:alphafe}
[\alpha/{\mbox{Fe}}] = 0.16 + 0.062(\log\mstar/\msun - 10).
\end{equation}
A major caveat of this renormalization is that it is based on
measurements of $\alpha$/Fe for {\em elliptical} galaxies, which are
predominately not forming stars, with an intermediate step of
connecting $\mstar$ to galaxy velocity dispersions.  Furthermore, the
\citet{gallazzi05} data are for both star-forming and
non--star-forming galaxies; in particular, the $\mstar\lesssim
10^{10.5}\,\msun$ steep part of the relation is predominately star
forming galaxies, while the flatter higher mass portion is
predominately passive galaxies.  At $\mstarnot\gtrsim
10^{11.5}\,\msun$, the predictions are not robust because of
extrapolation of Equation\,(\ref{eqn:fmr}) to very high star formation
rates.\footnote{Interestingly, there is increasing evidence that at
  high $\mstar$ the dependence in \tlogoh\ on star formation rate
  reverses \citep{yates12}.}

Overall, the agreement of our simple model predictions with the
observed average stellar metallicity relation is remarkably good over
$\sim$~three orders of magnitude.  Our prediction for the mean stellar
metallicity is in excellent agreement with the \citet{gallazzi05}
results at around the stellar mass of the Milky Way (few$\times
10^{10}\, \msun$), where they are expected to be the most robust. 
At stellar masses above $\sim 10^{11}\, \msun$, our predicted
metallicities turn down, while the \citet{gallazzi05} results continue
to increase---this is likely because of the breakdown of our
assumptions at these high masses, where the local galaxy population is
dominated by galaxies that are not on the star forming main sequence,
and which are likely to have experienced significant growth through
gas-poor mergers \citep{delucia07,skelton09}.

Between $10^{9} \lesssim \mstar \lesssim 10^{10}\, \msun$, our
predictions overshoot the \citeauthor{gallazzi05}\ results slightly
but are again in very good agreement with the results of \citet{woo08}
based on local group galaxies from $10^{8}$ to a few$\times 10^9\,
\msun$ \citep[a relation that extends four dex further down in
$\mstar$; ][]{kirby10,kirby11b}.  However, it is important to note
that the \citet{leitner12} star formation histories (and the
\citealt{mannucci10} metallicity relation) used in our predictions are
entirely extrapolated in this regime. At $\mstar\sim 10^8$, the
\citeauthor{leitner12} extrapolated star formation histories differ
substantially from the histories derived from the ``fossil record''
(resolved color magnitude data for individual stars) in nearby dwarf
galaxies \citep{weisz11}. The Leitner estimates would imply that dwarf
galaxies form most of their mass at $z<1$, while observationally,
downsizing reverses in this regime: the \citet{weisz11} results imply
that they form most of their stars at $z>2$ (see \citealt{leitner12}
for a detailed discussion).  In our framework, however, $z=0$
metallicities do not depend on {\em when} the metals were made: as
long as the ensemble-averaged galaxies remain on the
$\mstar$-$\msfr$-$\zg$ relation, the average stellar metallicities
will be well-described by our model.\footnote{Furthermore, it is
  intriguing to note that the observed average stellar metallicities
  in this mass range obey such a tight relation despite the wide range
  in metallicity distribution functions \citep{dolphin05} and the
  relatively wide spread in gas-phase abundances
  \citep[e.g.,][]{zahid12a}.}

Thus, at both low and high masses, it is difficult to compare our
predictions to the current observations.  Based on this combination of
observational factors and the regimes in which we are extrapolating to
get predictions, we consider our predictions to be the most robust at
stellar masses roughly comparable to the Milky Way ($\mstarnot\sim 1$
to $5\times 10^{10}\,\msun$).

Finally, the steep drop in predicted metallicities at
$\log\mstar/\msun\lesssim 9.5$ is likely caused by an underestimate of
\tlogoh\ from extrapolating Equation\,(\ref{eqn:fmr}) outside the
observed range of $\mstar$\ and $\msfr$.  Though the fit given by
\citeauthor{laralopez10} (equation\,\ref{eqn:lara}) gives nearly the
same $\alphastar$ at low $\mstar$ as shown in
Figure\,\ref{fig:stellar}, the piecewise-linear fit from
\citeauthor{mannucci10}\ (their equation~5) gives higher metallicities
for low-mass galaxies ($\log\alphastar/\alpha_{\odot}\sim -0.5$ at
$\mstar\sim10^8\msun$), though it dramatically over-predicts the
metallicities of high-mass galaxies.

\begin{table}
\centering
\begin{tabular}{rr|rr|rr}\hline\hline
\multicolumn{2}{c}{$z=1.1$} & \multicolumn{2}{c}{$z=0.7$} & \multicolumn{2}{c}{$z=0.0$} \\
$\log\mstar$ & $\log\alphastar$ & $\log\mstar$ & $\log\alphastar$ & $\log\mstar$ & $\log\alphastar$ \\ \hline
--- & --- & --- & --- & $8.00$ & $-0.93$ \\ 
--- & --- & --- & --- & $8.50$ & $-0.55$ \\ 
--- & --- & $7.99$ & $-1.20$ & $9.00$ & $-0.25$ \\ 
--- & --- & $8.53$ & $-0.75$ & $9.25$ & $-0.12$ \\ 
$7.72$ & $-1.59$ & $8.97$ & $-0.45$ & $9.50$ & $-0.01$ \\ 
$8.62$ & $-0.79$ & $9.35$ & $-0.23$ & $9.75$ & $ 0.08$ \\ 
$9.22$ & $-0.41$ & $9.70$ & $-0.07$ & $10.00$ & $ 0.16$ \\ 
$9.68$ & $-0.17$ & $10.02$ & $ 0.05$ & $10.25$ & $ 0.22$ \\ 
$10.08$ & $-0.01$ & $10.33$ & $ 0.13$ & $10.50$ & $ 0.26$ \\ 
$10.44$ & $0.09$ & $10.62$ & $0.19$ & $10.75$ & $0.29$ \\
$10.77$ & $ 0.15$ & $10.90$ & $ 0.22$ & $11.00$ & $ 0.29$ \\
$11.07$ & $ 0.18$ & $11.18$ & $ 0.23$ & $11.25$ & $ 0.28$ \\ 
$11.37$ & $ 0.18$ & $11.45$ & $ 0.22$ & $11.50$ & $ 0.26$ \\
\hline
\end{tabular}
\caption{\label{tbl:mzevolve} 
Evolution of the stellar metallicity-mass relation, as plotted in the
bottom panel of Figure\,\ref{fig:stellar}; stellar masses $\mstar$ are
in units of $\msun$ and metallicities $\alphastar$ are relative to
$\alpha_{\odot}$. Rows denote the growth of individual galaxies.
}
\end{table}

The bottom panel of Figure\,\ref{fig:stellar} and
Table\,\ref{tbl:mzevolve}\ give the predicted evolution of the
$\alphastar$-$\mstar$ relation from $z=1.1$, $0.7$, to
$0.0$.\footnote{Though stellar metallicities have been measured at
  redshifts as high as $z\sim 3$ \citep{sommariva12}, we do not
  provide predictions for stellar metallicities at higher redshift due
  to uncertainties in the star formation histories.}  The stellar
metallicity-mass relation is slightly steeper at higher redshift than
at $z=0$; this evolution is driven by the relatively late buildup of
stellar mass in low-mass galaxies relative to more massive galaxies,
as can be seen in the top panel of Figure\,\ref{fig:scaling}.  In
particular, at $\mstar\gtrsim 10^{10.5}\,\msun$, where ongoing surveys
are measuring stellar metallicities at $z\sim 0.7$ for the first time
(Gallazzi et al.\ in preparation), we find a typical decrease in
$\alphastar$ of less than $\sim 0.1$\,dex for star-forming galaxies.
(Note that once star formation has ceased, stellar metallicities will
not evolve; thus if passive galaxies at $z=0$ have not formed stars
since $z=0.7$,, then galaxies will either remain in place or move to
the right in the $\zstar$-$\mstar$ plane through stellar mass growth
via dry mergers).  This rather modest evolution is in stark contrast
to the evolution seen at all stellar masses in the $\zg$-$\mstar$
relation (top-right panel of Figure~\,\ref{fig:scaling}); this
difference is primarily due to the earlier buildup of stellar mass in
the more massive galaxies.

\section{Conclusions}\label{sec:conc}
We have combined empirically derived star-formation histories from
observations of high redshift galaxies \citep{leitner12} with the
empirically determined relation between stellar mass, star formation
rate, and gas-phase metallicity \citep{mannucci10} to derive the
distribution of stellar metallicities at $z=0$ and the evolution of
the galaxy-averaged stellar metallicity-mass relation
(Figure\,\ref{fig:stellar} and Tables\,\ref{tbl:hist} and
\ref{tbl:mzevolve}).  We find that the hypothesis that the observed
$z=0$ relation between stellar mass, star formation rate, and
gas-phase metallicity holds up to at least $z\sim2.5$ is consistent
with stellar metallicities, within the limits of the current
observations.  This comparison, however, is hampered by both a lack of
a well-measured $\mstar$-$\msfr$-$\zg$ relation at low stellar masses,
low star formation rates, and high star formation rates, and a lack of
well-characterized $\alpha$-element distribution functions in both the
Milky Way and other star-forming field galaxies with
$10^8\lesssim\mstar\lesssim 10^{11}\msun$. From the modeling side,
upcoming characterizations of the $\mstar$-$\msfr$-$\zg$ relation to
lower stellar masses and across a broader range of star formation
rates (B.\ Andrews \& P.\ Martini, in preparation) will help alleviate
the need for extrapolation when calculating \tlogoh.  Upcoming surveys
will also provide a more complete census of the stellar metallicity
distribution function for the Milky Way (e.g., the SDSS-III APOGEE
survey) and Local Group dwarf galaxies (E.\ Kirby et al.\ in
preparation).  

We have assumed here that all of the scatter in \tlogoh\ at fixed
stellar mass is related to variations in the star formation rate, but
the resulting relation still has scatter that is correlated with,
e.g., galaxy size \citep{ellison08a,yabe12} or environment
\citep{kewley06a,ellison08b,pasquali12,hughes12}.  Moreover, the
scatter in $\zstar$ at fixed stellar mass is known to correlate with
galaxy age \citep{gallazzi05}.  In our framework, this can be seen as
variations in galaxies' star formation histories via the
$\sim0.3$\,dex\ scatter in star formation rate at fixed stellar mass
\citep{noeske07a,karim11,leitner12}.  This variation can be partially
attributed to changes in environment \citep{pasquali10}, especially if
environment introduces additional scatter to the
$\zg$-$\mstar$-$\msfr$\ relation \citep[e.g.,][]{pasquali12}.

We have also assumed that the buildup of {\em all} of the stellar mass
in a galaxy can be well described by Equation\,(\ref{eqn:sfms}).  This
is clearly not the case, as, e.g., there are stars in the Milky Way
that were formed before $z\sim2.5$ and with $\log \zstar/\zsun < -3$.
However, these stars comprise a very small fraction of the total
stellar mass or light, at least in the Milky Way, and could be
addressed by more complete model of star formation histories,
including minor mergers which bring in a populations of
lower-metallicity stars.

One of the strengths of the method presented here is that it does not
require us to explain {\em why} the relationship between gas-phase
metallicity, stellar mass, and star formation rate apparently does not
evolve with redshift by invoking carefully evolving gas fractions,
accretion rates, or outflow efficiencies \citep[such as in,
e.g.,][]{peeples11,dave12,dayal12}.  On the other hand, integrating
over the evolution of the \mzr\ does neatly predict what fraction of
Oxygen ever produced by a $z=0$ galaxy should be still locked up in
stars, and how steeply this fraction should increase with stellar mass
\citep[c.f.,][]{gallazzi08,kirby11c,zahid12b}.  Moreover, by combining
this kind of calculation with measurements of the \mzr\ and $z=0$\ gas
fractions \citep[e.g.,][]{peeples11}, one can place constraints on the
total amount of Oxygen galaxies have expelled over their lifetime
\citep{bouche07,kirby11c,zahid12b}, though for star-forming galaxies,
uncertainties in the gas-phase abundances and relevant gas masses make
this calculation difficult. More globally, it is especially intriguing
that the implied masses of both metals and all baryons expelled by
galaxies over their lifetime appears potentially consistent with the
oxygen mass observed in the circumgalactic medium of $z\sim 0.25$
galaxies (\citealp{tumlinson11}; J.\ Werk et al.\ in preparation).

\section*{Acknowledgments}
We are grateful to Sam Leitner for sharing his \citet{leitner12}\
star-forming main-sequence integration code with us.  We thank Andrea
Ferrara, Joseph Mu\~{n}oz, Evan Kirby, and Paul Martini for useful
discussions.  We are also appreciative for the exquisite comments on
the text from the anonymous referee. MSP acknowledges support from the
Southern California Center for Galaxy Evolution, a multi-campus
research program funded by the University of California Office of
Research.  This research was supported in part by the National Science
Foundation under Grant No. NSF PHY11-25915 as part of the Kavli
Institute for Theoretical Physics program on First Galaxies and Faint
Dwarfs.

\label{lastpage}
\end{document}